\documentclass{PoS}

\title{QCD phase diagram from the lattice at strong coupling}

\ShortTitle{QCD phase diagram from the lattice at strong coupling}

\author{Philippe de Forcrand\\
Institute for Theoretical Physics, ETH Z\"urich, CH-8093 Z\"urich, Switzerland and\\
CERN, Physics Department, TH Unit, CH-1211 Geneva 23, Switzerland\\
        E-mail: \email{forcrand@phys.ethz.ch}}

\author{Owe Philipsen, \speaker{Wolfgang Unger}\\
Institut f\"ur Theoretische Physik, Goethe-Universit\"at Frankfurt,\\
60438 Frankfurt am Main, Germany\\
        E-mail: \email{philipsen@th.physik.uni-frankfurt.de}\\
	\hspace{11mm} \email{unger@th.physik.uni-frankfurt.de}}

\abstract{
\vspace*{-12.0cm}
\begin{flushright}
\texttt{\footnotesize \textbf{CERN-PH-TH-2015-067}}\\
\end{flushright}
\vspace*{11.0cm}
The phase diagram of lattice QCD in the strong coupling limit can be measured in the full \linebreak $\mu$-$T$ plane, also in the chiral limit.
In particular, the phase diagram in the chiral limit features a tricritical point at some $(\mu_c,T_c)$.
This point may be related to the critical end point expected in the QCD phase diagram.
We discuss the gauge corrections to the phase diagram at strong coupling and compare our findings with various possible scenarios in continuum QCD. We comment on the possibility that the
tricritical point at strong coupling is connected to the tricritical point in the continuum, massless QCD.
}

\FullConference{9th International Workshop on Critical Point and Onset of
Deconfinement\\
                 17-21 November, 2014\\
                 ZiF (Center of Interdisciplinary Research), University of Bielefeld, Germany}

\usepackage{amsmath}
\usepackage{amssymb}
\usepackage{bbm}
\usepackage{caption}

\def \Ns {{N_{\sigma}}}
\def \Nt {{N_{\tau}}}
\def \Nc {3}
\def \Nf {{N_{f}}}

\newcommand{\expval}[1]{\left\langle #1 \right\rangle}

\newcommand{\pbp}{\bar{\psi}\psi}

\newcommand{\lr}[1]{\left( #1 \right)}

\newcommand{\beqn} {\begin{equation}}
\newcommand{\eqn} {\end{equation}}

\def \beq{\begin{equation}}
\def \eeq{\end{equation}}
\def \bea{\begin{eqnarray}}
\def \eea{\end{eqnarray}}

\def \Tr {{\rm Tr}}

\def \bet0{\beta_0}
\def \bet1{\beta_1}
\def \simgt{\,\rlap{\lower 7.5 pt\hbox{$\mathchar \sim$}}\raise 3 pt \hbox{$>$}\,}
\def \simlt{\,\rlap{\lower 7.5 pt\hbox{$\mathchar \sim$}}\raise 3 pt \hbox{$<$}\,}

\def\lsim{\raise0.3ex\hbox{$<$\kern-0.75em\raise-1.1ex\hbox{$\sim$}}}
\def\gsim{\raise0.3ex\hbox{$>$\kern-0.75em\raise-1.1ex\hbox{$\sim$}}}

\newcommand{\hnu}{{\hat{\nu}}}

\begin{document}

\section{Motivation}

The QCD phase diagram is conjectured to have a rich phase structure. At low temperatures, QCD has a vacuum and nuclear matter phase; at high temperatures and/or densities, 
QCD matter develops a qualitatively different phase where quarks are liberated from confinement - the so-called quark gluon plasma (QGP).
While there is strong evidence for a crossover transition from the hadronic phase to the QGP for zero baryon chemical potential $\mu_B$, there is no evidence for a true phase transition at higher densities.
Lattice studies of QCD have aimed to extend the simulations to finite quark chemical potential $\mu=\frac{1}{3}\mu_B$, but the available methods are limited to $\mu/T \lesssim 1$ due to the sign problem: 
Monte Carlo simulations sample a probability distribution and hence rely on the condition that the statistical weights are positive. 
In the conventional approach to lattice QCD based on the fermion determinant, the weight for the fermion determinant becomes complex as soon as the chemical potential is non-zero. 
The sign problem (more precisely in this context: the complex phase problem) is severe, prohibiting direct simulations for $\mu>0$ - which is also due to the fact that Monte Carlo is performed on the colored gauge fields. 

However, there is a representation of lattice QCD which does not suffer severely from the sign problem: in this representation, the lattice degrees of freedom are color singlets.
The complex phase problem is reduced to a mild sign problem induced by geometry-dependent signs of fermionic world lines.
Such a ``dual'' representation of lattice QCD has been derived for staggered fermions in the strong coupling limit, that is in the limit of infinite gauge coupling $g\rightarrow \infty$ \cite{Rossi1984}.
In this limit, only the fermionic action contributes to the path integral, whereas the action describing gluon propagation is neglected.
QCD at strong coupling has been studied extensively since 30 years, both with mean field methods~\cite{Kawamoto1981,Damgaard1985,Bilic1992a,Bilic1992b,Nishida2004,Ohnishi2009} 
and by Monte Carlo simulations~\cite{Rossi1984,Wolff1985,Karsch1989,Adams2003,Fromm2010,Unger2011}. 
Those studies have been limited to the strong coupling limit, which corresponds to rather coarse lattices.
However, recently \cite{Unger2014} we were able to include the leading order gauge corrections to the partition function.
The effects of these gauge corrections on the phase diagram will be discussed below.

\section{The chiral and nuclear transition in the strong coupling limit}

The path integral of staggered fermions in the strong coupling limit can be rewritten exactly as a
partition function of a monomer+dimer+flux system.
The reformulation proceeds in two steps: first the gauge links (gluons) are integrated out, 
which confines the quark fields $\psi(x)$ into color singlets, the hadrons: those are the mesons $M(x)=\bar{\psi}(x)\psi(x)$ and the baryons\linebreak
$B(x)=\frac{1}{6}\epsilon_{i_1 i_2 i_3}\psi_{i_1}(x)\psi_{i_2}\psi_{i_3}(x)$.
In the second step, also the	 quarks are integrated out, which allows to express the partition function via integer variables:
\begin{align}
Z_{SC}(m_q,\mu)= \sum_{\{k_b,n_x,\ell\}}
\underbrace{\prod_{b=(x,\mu)}\frac{(\Nc-k_b)!}{\Nc!k_b!}}_{\text{meson hoppings}\, M_xM_y}
\underbrace{\prod_{x}\frac{\Nc!}{n_x!}(2am_q)^{n_x}}_{
\text{chiral condensate}\, M_x}
\underbrace{\prod_\ell w(\ell,\mu)}_{\text{baryon hoppings}\, \bar{B}_xB_y}
\label{SCPF}
\end{align}
with $k_b\in \{0,\ldots 3\}$, $n_x \in \{0,\ldots 3\}$, $\ell_b \in \{0,\pm 1\}$.
Since the quark fields are treated as anti-commuting Grassmann variables in the path integral, the integration realizes a Pauli exclusion principle called the Grassmann constraint:
\begin{equation}
n_x+\sum_{\hnu=\pm\hat{0},\ldots, \pm \hat{d}}\lr{k_{\hnu}(x) + \frac{N_c}{2} |\ell_\hnu(x)|} = 3.
\label{Grassmann}
\end{equation}
This constraint restricts the number of admissible configurations ${\{k_b,n_x,\ell\}}$ in Eq.~(\ref{SCPF}) such that mesonic degrees of freedom always add up to $\Nc$ and baryons form self-avoiding loops not in contact with
the mesons. The weight $w(\ell,\mu)$ and sign $\sigma(\ell)=\pm 1$ for an oriented baryonic loop $\ell$ depend on the loop geometry.
The partition function Eq.~(\ref{SCPF}) describes effectively only one quark flavor, which however corresponds to four flavors in the continuum (see Sec.~\ref{NfDep}). It is valid for any quark mass. We will however restrict here to the theoretically most interesting case of massless quarks, $m_q=0$.
In fact, in this representation the chiral limit is very cheap to study via Monte Carlo, in contrast to conventional determinant-based lattice QCD where the chiral limit is prohibitively 
expensive.

\begin{figure}
\raisebox{-3mm}{
\includegraphics[width=0.47\textwidth]{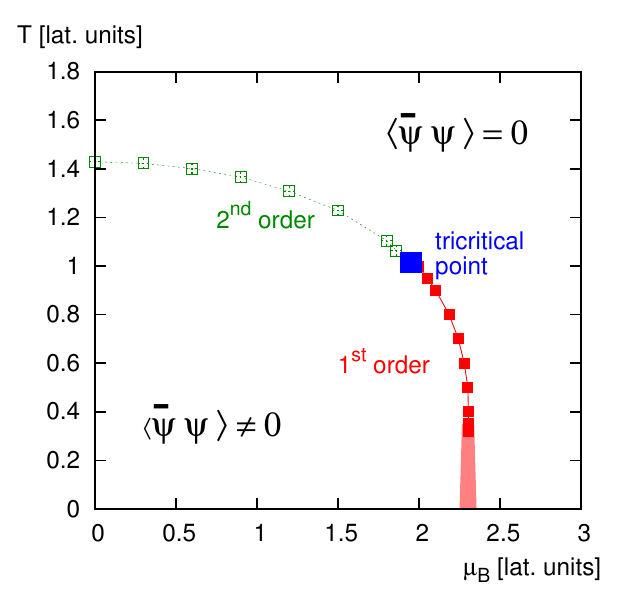}}
\includegraphics[width=0.49\textwidth]{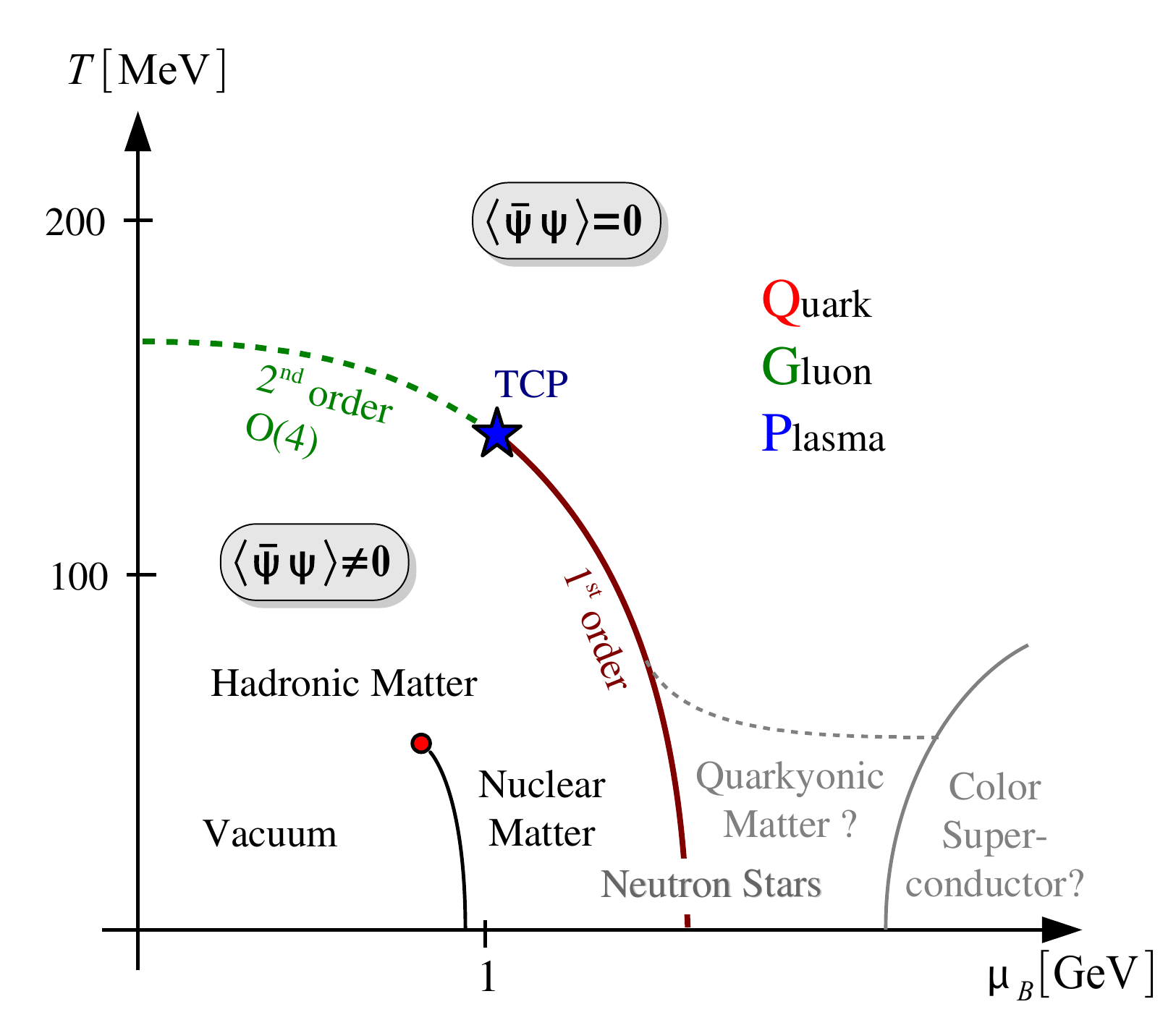}
\caption{The Phase diagram in the strong coupling limit (\emph{left}), as measured in a Monte Carlo simulation, compared to the standard expectation of the continuum QCD phase diagram (\emph{right}). Both diagrams are for massless quarks.}
\label{PhaseDiagCL}
\end{figure}

For staggered fermions in the strong coupling limit, there is a remnant of the chiral symmetry $U_{55}(1)\subset SU_L(\Nf)\times SU_R(N_f)$.
This symmetry is spontaneously broken at $T=0$ and is restored at some critical temperature $T_c$ with the chiral condensate $\expval{\pbp}$ being the order parameter of this transition.
As shown in Fig.~\ref{PhaseDiagCL} (\emph{left}), we find that this transition is of second order. This is analogous to the standard expectation in continuum QCD with $\Nf=2$ massless quarks,
where the transition is also believed to be of second order. Moreover, both for our numeric finding at strong coupling and for the expectation in the continuum, the transition turns into first order as the chemical potential is increased. 
Thus the first order line ends in a tricritical point, which is the massless analogue of the chiral critical endpoint sought for in heavy ion collisions.

In fact, at strong coupling, the zero temperature nuclear transition at $\mu_{B,c}\simeq m_B$ is intimately connected to the chiral transition, and they coincide as long as the transition is first order.
The reason for this is the saturation on the lattice due to the Pauli principle: in the nuclear matter phase at $T=0$, the lattice is completely filled with baryons, leaving no space for a non-zero chiral condensate to form 
(in terms of the dual variables, there is no space for monomers on the lattice).
This is certainly a lattice artifact which disappears in the continuum limit, where the nuclear phase behaves like a liquid rather than a crystal.

The ultimate question is whether the tricritical point at strong coupling is related to the hypothetical tricritical point in continuum massless QCD. If we can establish such a connection numerically, 
this would be strong evidence for the existence of a chiral critical endpoint in the $\mu$-$T$ phase diagram of QCD. To answer this question, it is necessary to go away from the strong coupling limit and incorporate the 
gauge corrections, which will lower the lattice spacing and eventually allow to make contact to the continuum.

\section{Gauge Corrections to the strong coupling phase diagram}

\begin{figure}
\centerline{
\includegraphics[width=0.9\textwidth]{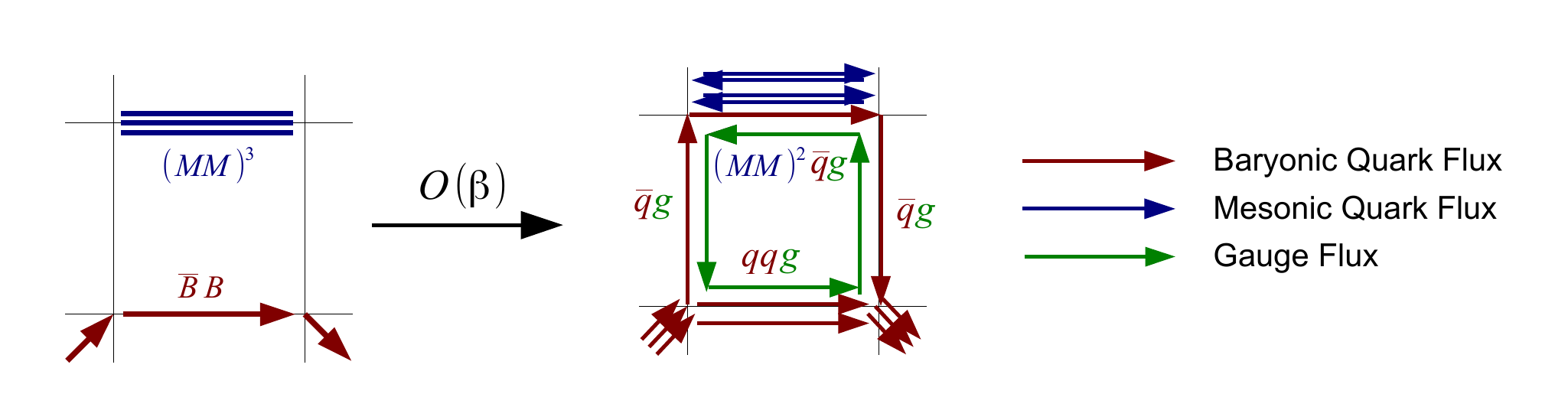}
}
\caption{Example of an $\mathcal{O}(\beta)$ diagram. On an excited plaquette, color singlets can also be composed of quark-quark-gluon or antiquark-gluon combinations. Whereas in the strong coupling limit baryons are pointlike,
they become extended objects due to the gauge corrections.}
\label{weight}
\end{figure}
Lattice QCD in the strong coupling limit is defined by the the lattice coupling $\beta=\frac{6}{g^2}\rightarrow 0$ as $g\rightarrow \infty$.
Going away from the strong coupling limit is realized by making use of strong coupling expansions in $\beta$.
We have recently shown how to incorporate the leading order gauge corrections $\mathcal{O}(\beta)$ \cite{Unger2014}.
In a nutshell, the strategy is to compute link integrals at the boundary of ``excited'' plaquettes, which correspond to gluonic excitations.
Introducing a variable $q_P \in \{0,1\}$ to mark the ''excited'' plaquettes $P$, the ${\cal O}(\beta)$ partition
function can be expressed in a similar fashion as Eq.~(\ref{SCPF}) with modified weights $\hat{w}$ (for details see \cite{Unger2014}):
\begin{align}
Z(\beta) &=\sum_{\{n,k,\ell,q_P\}}\prod_x \hat{w}_x \prod_{b} \hat{w}_b \prod_\ell \hat{w}_\ell \prod_P \hat{w}_P,\qquad \hat{w}_P=\left(\frac{\beta}{6}\right)^{q_P}.
\end{align}
We can sample this partition function by the same algorithm (variant of the worm algorithm) as for $\beta\!=\!0$, adding a Metropolis accept/reject
step to update the plaquette variables $q_P$.
These simulations have been carried out for $\Nt=4$ and various lattice volumes $\Ns=4,6,8,12,16$ to perform finite size scaling and to measure the phase boundary as a function of the chemical potential.
In contrast to the strong coupling limit, where the color singlets are entirely composed of quarks and anti-quarks, including the gauge corrections allows color singlets to be composed
of quark-quark-gluon or antiquark-gluon color singlet states, as shown in Fig.~\ref{weight}.
There are two qualitatively new features that arise when incorporating the $\mathcal{O}(\beta)$ corrections:

\begin{figure}[t!]
\includegraphics[width=\textwidth]{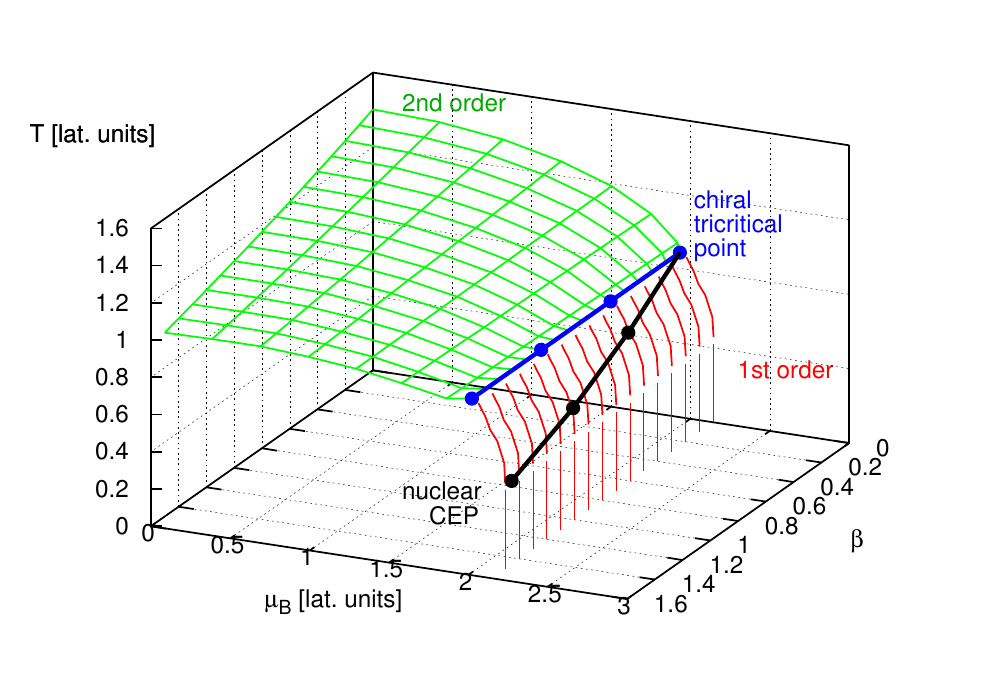}
\caption{
Phase boundary in the $\mu$-$T$ plane extended to finite $\beta$. The backplane corresponds to the strong coupling limit $\beta=0$. The second order phase boundary is lowered by increasing $\beta$.
We do not observe a shift of the chiral tricritical point.
However, the nuclear critical endpoint (CEP), determined from 
the baryon density, moves down along the
first order line (extrapolated to $T=0$ to guide the eye) as $\beta$ is increased.
}
\label{finiteMu}
\end{figure}

\begin{enumerate}
 \item Baryons are point-like in the strong coupling limit, the lattice spacing is too coarse to resolve the internal structure of the baryon. Including the gauge correction, baryons become extended objects, 
 spread over one lattice spacing.
 \item The nuclear potential in the strong coupling limit is of entropic nature, where two static baryons interact merely by the modification of the pion bath. With the leading order gauge correction, 
pion exchange is possible as the Grassmann constraint is relaxed: on excited plaquettes, the degrees of freedom in Eq.~(\ref{Grassmann}) add up to $4$ instead of $3$.
\end{enumerate}

These features will have an impact on the phase boundary. In Fig.~\ref{finiteMu}, the effect of the gauge corrections is shown. We find that the second order phase boundary is lowered, as expected because 
the critical temperature in lattice units drops as the lattice spacing is decreased with increasing $\beta$. However, we find the chiral tricritical point and the first order transition to be invariant under 
the $\mathcal{O}(\beta)$ corrections. We want to stress that there are actually two end points, which split due to the gauge corrections:
the second order end point of the nuclear liquid-gas transition is traced by looking at the nuclear density as an order parameter. 
We expect the nuclear and the chiral first order transition to split, such that at $T=0$ there are three different phases instead of two phases (as shown in Fig.~\ref{PhaseDiagCL} \emph{right}).
The nuclear phase is in the continuum distinct from the chirally restored phase. As a first evidence for this splitting, we find that the nuclear critical end point separates from the chiral tricritical point.

\section{Relation between the strong coupling phase diagram and continuum QCD}
\label{NfDep}

\begin{figure}
\vspace{-5mm}
\centerline{(A)\hspace{4.5cm} (B)\hspace{4.5cm} (C)}
\centerline{\includegraphics[width=0.35\textwidth]{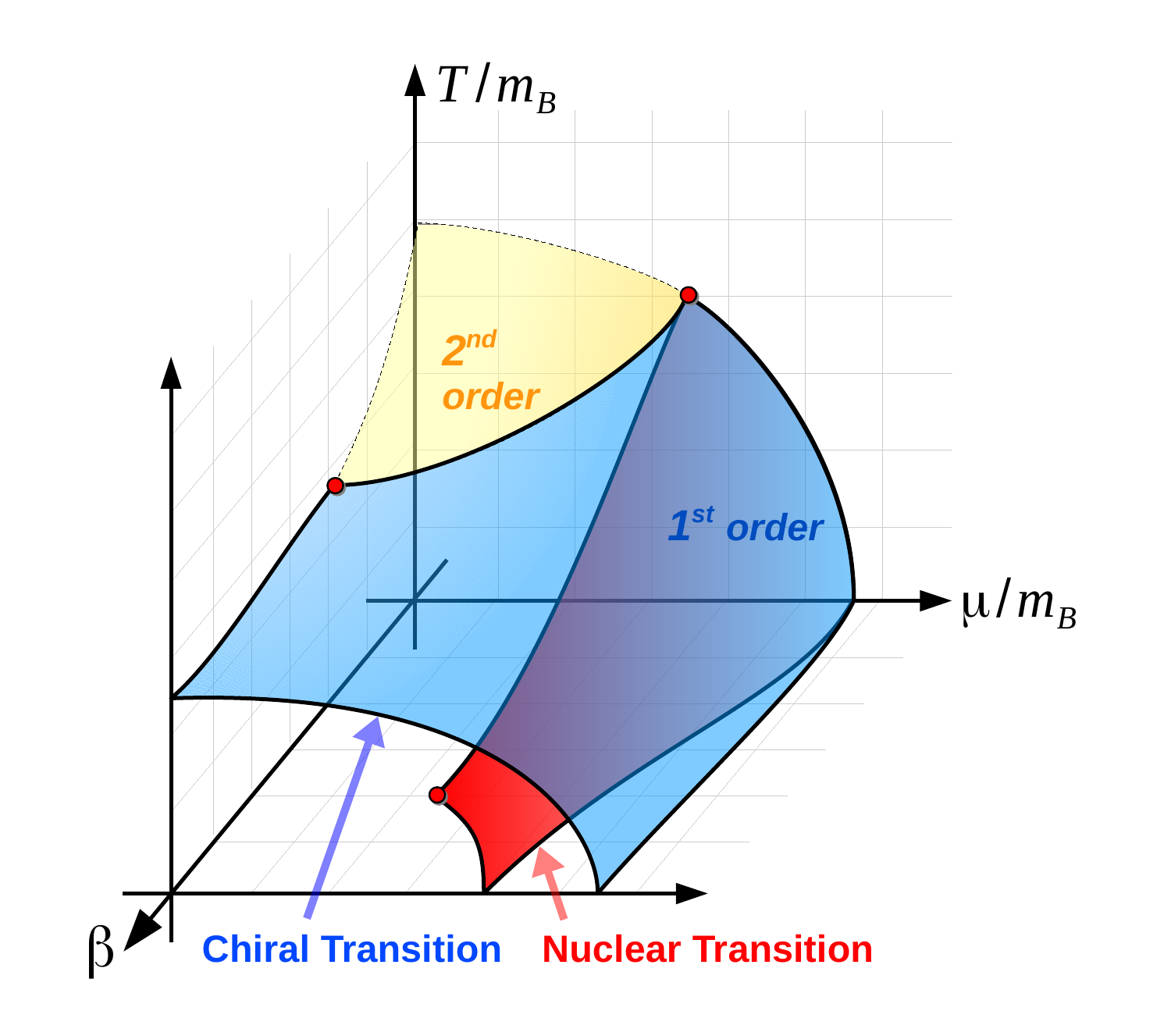}\hspace{-3mm}
\includegraphics[width=0.35\textwidth]{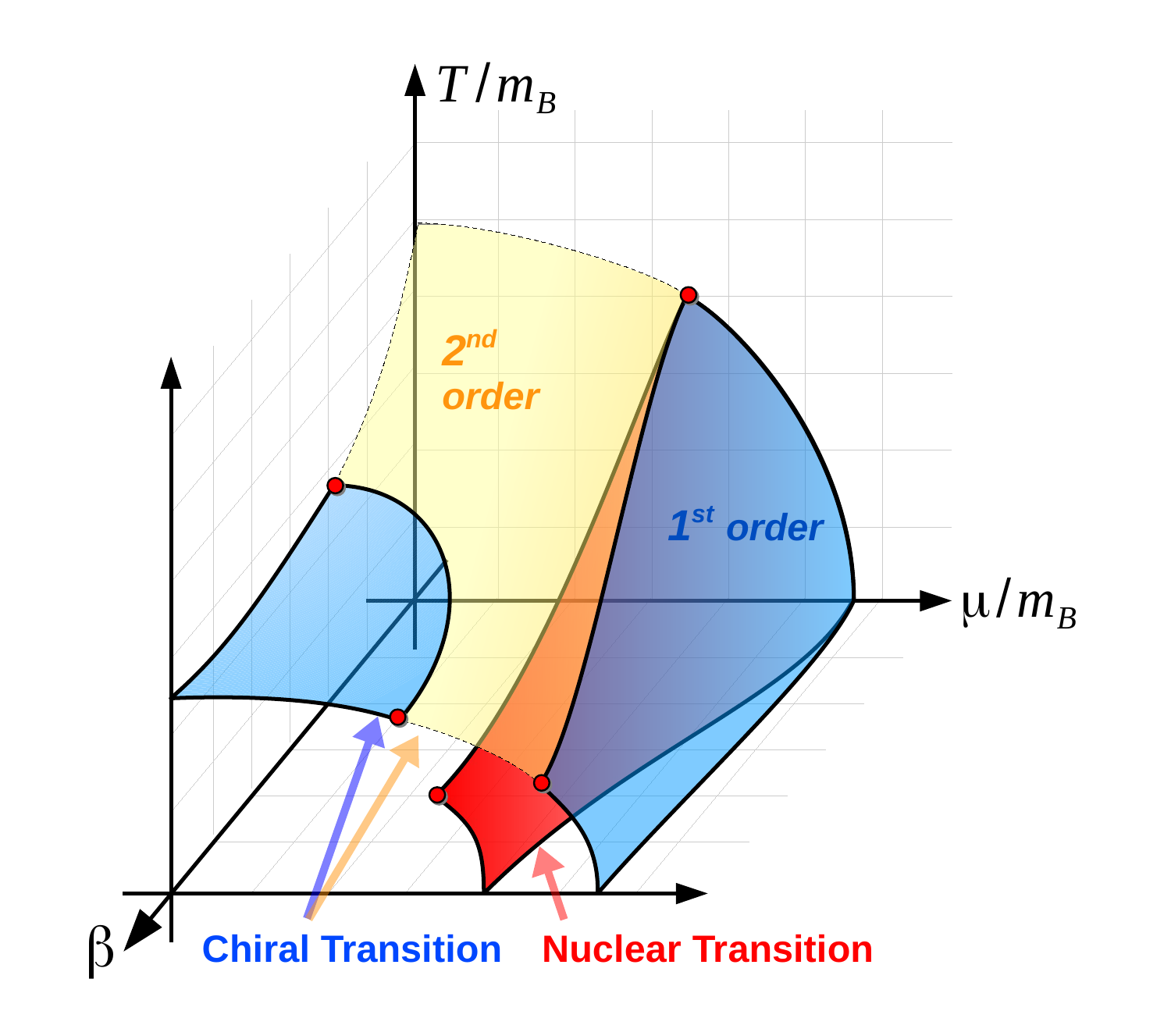}\hspace{-3mm}
\includegraphics[width=0.35\textwidth]{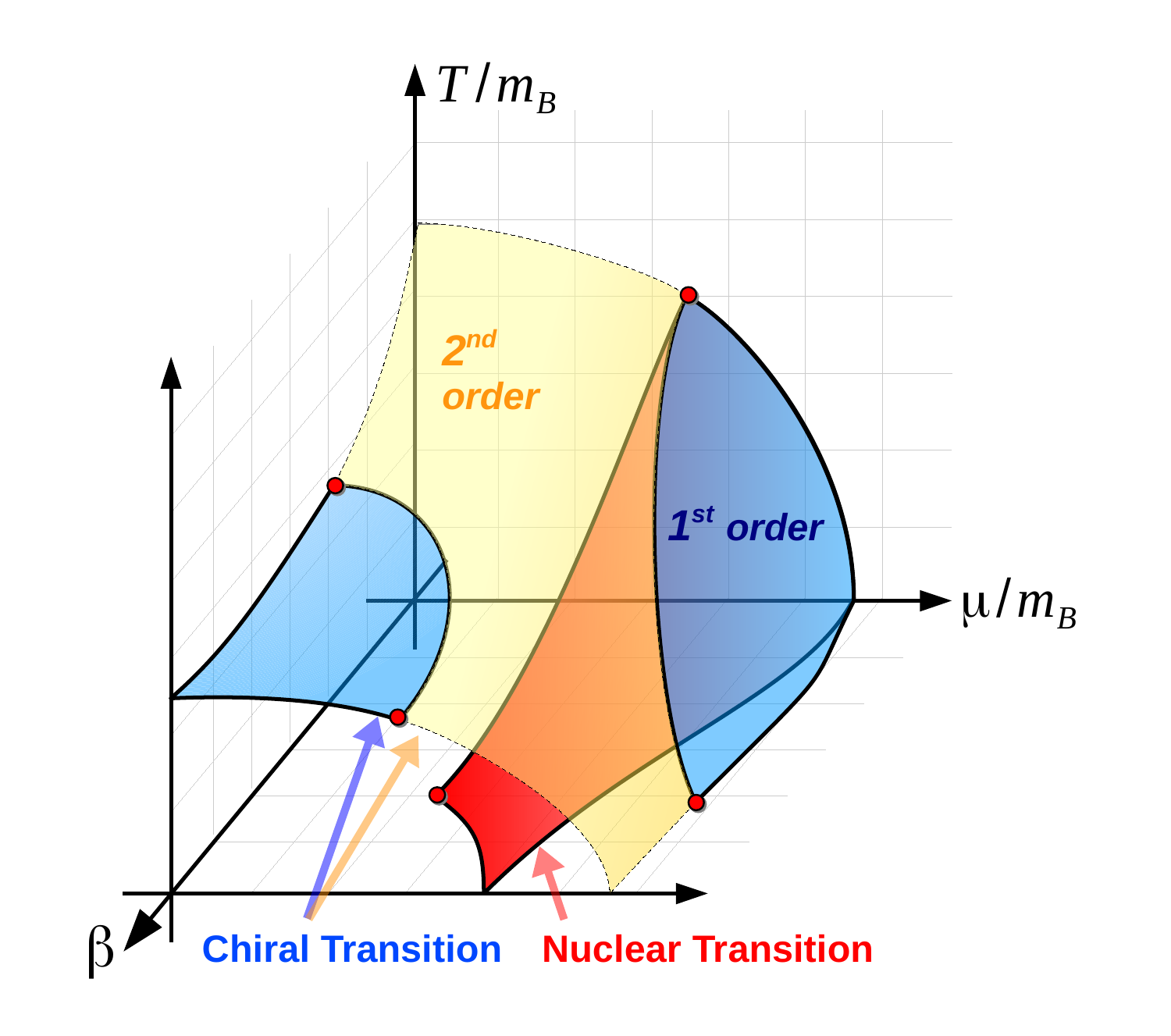}}
\centerline{\includegraphics[width=0.35\textwidth]{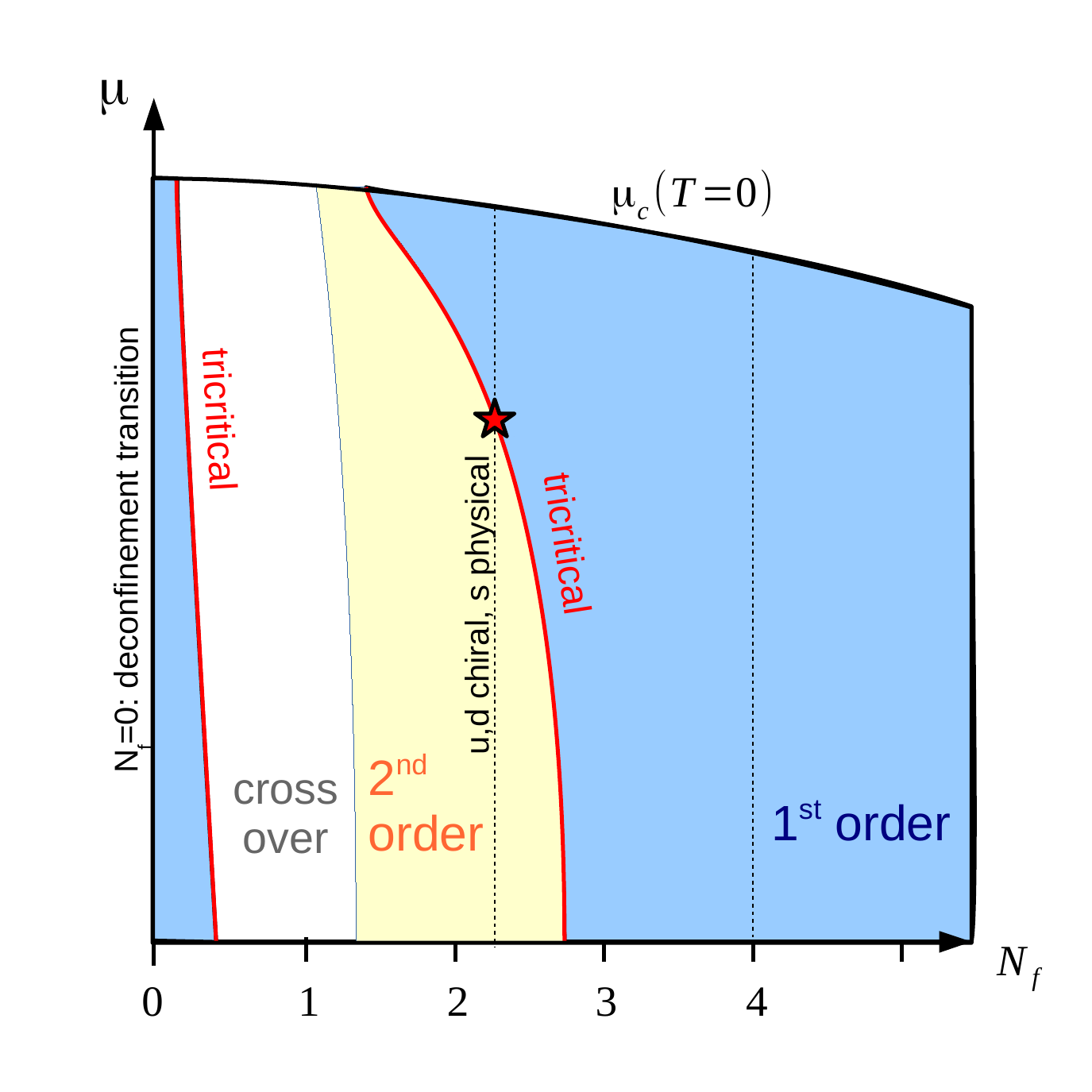}\hspace{-3mm}
\includegraphics[width=0.35\textwidth]{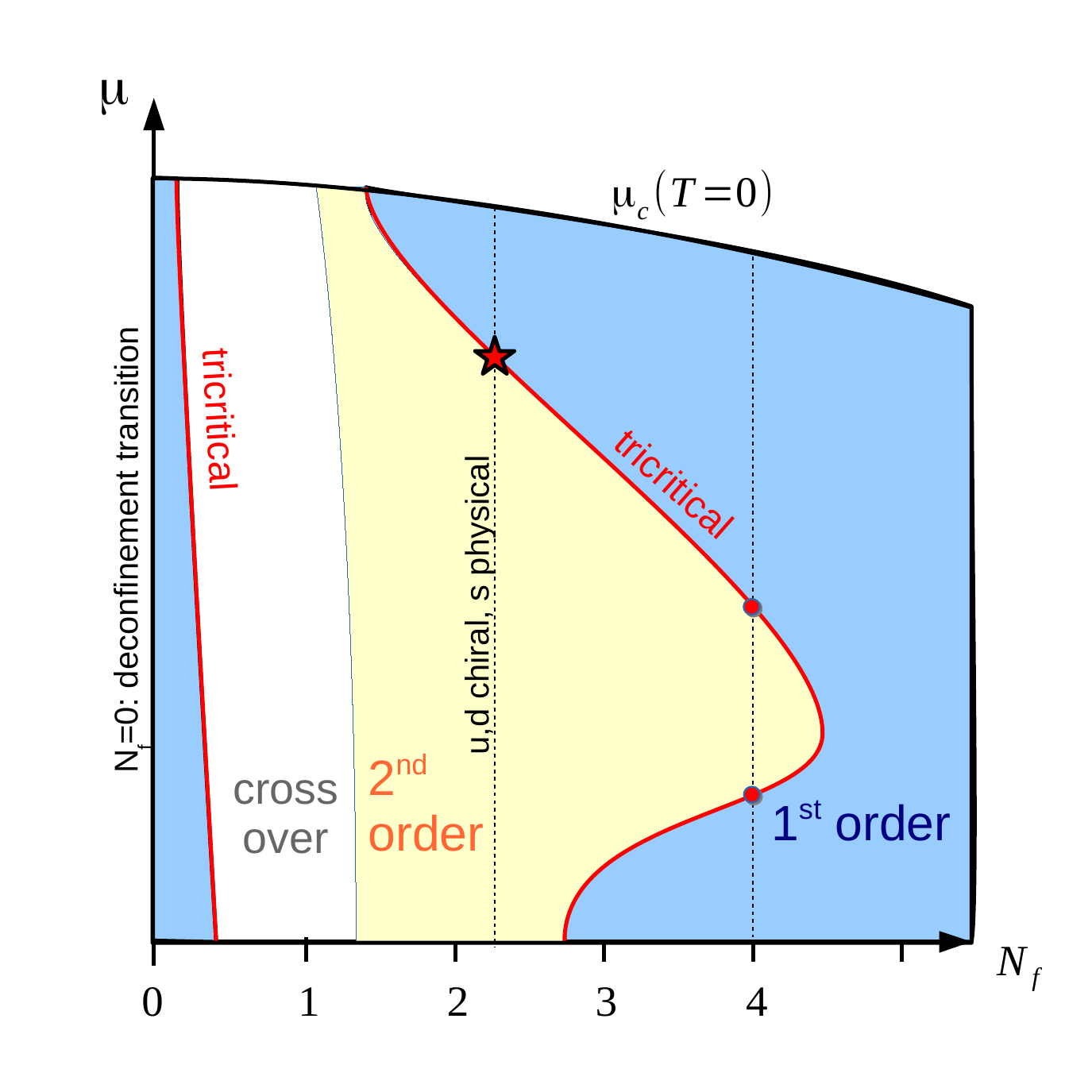}\hspace{-3mm}
\includegraphics[width=0.35\textwidth]{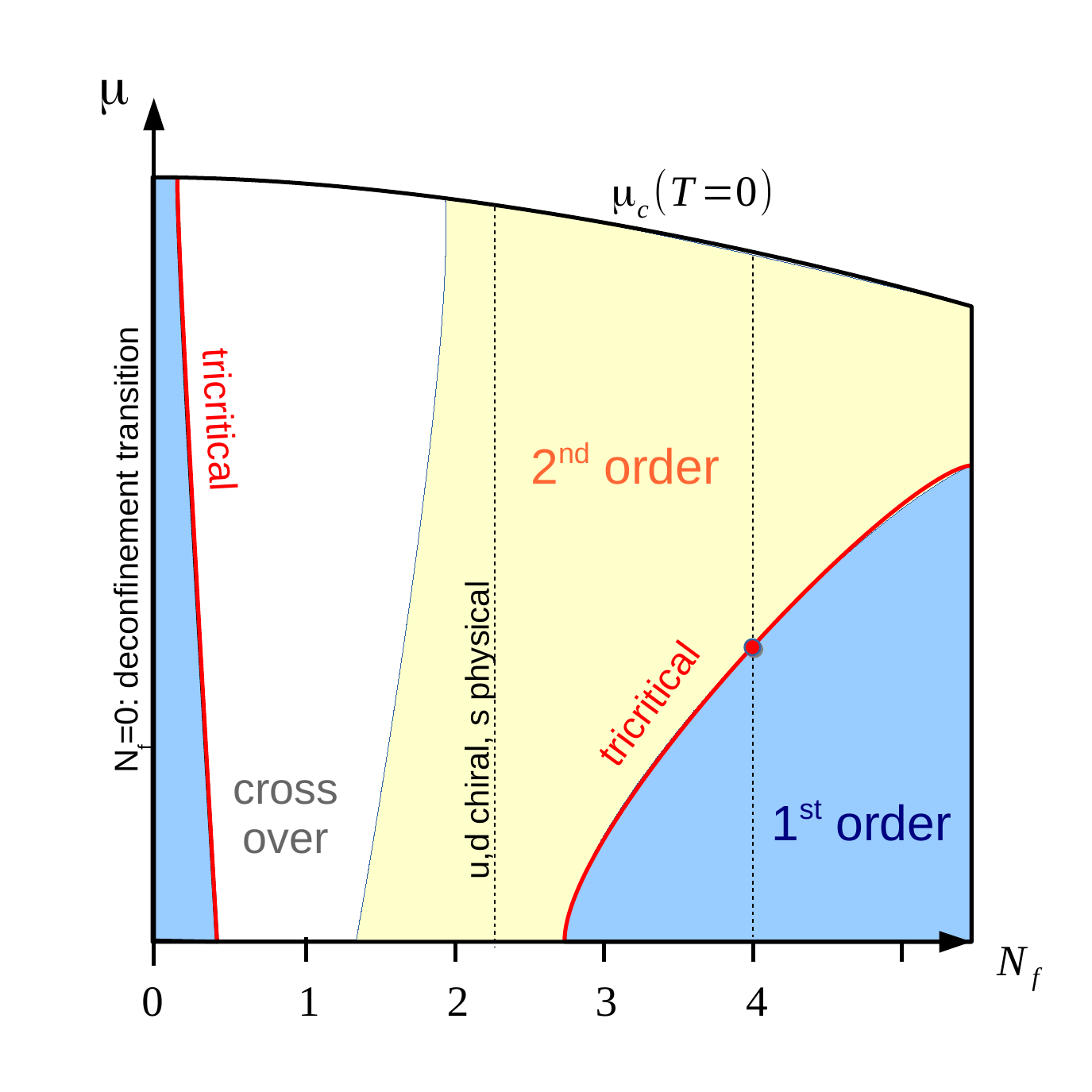}}
\caption{
\emph{Top row:} Various scenarios for extending the phase diagram in the strong coupling limit ($\beta=0$) toward the the continuum limit ($\beta\rightarrow \infty$).
All three scenarios assume that the nuclear and chiral transition split, and that at $\mu=0$ the chiral transition is of first order (since in the continuum $\Nf=4$).
In the strong coupling limit, the chiral transition at $\mu=0$ is second order (corresponding to $\Nf=1$ and the doublers decoupled), hence there must be a tricritical point
at some $\beta_{tric}^{(\mu=0)}$. It is an open question whether the tricritical point at strong coupling is connected to this tricritical point at $\beta_{tric}^{(\mu=0)}$ (\emph{left}), or connected to the speculated 
tricritical point in the continuum (\emph{center}) or terminates at some finite $\beta$ at $T=0$ (\emph{right}).\\ 
\emph{Bottom row:} the corresponding scenarios for the finite temperature chiral transition in the $\mu-\Nf$ phase diagram, showing the possible relation of the tricritical point at $\Nf=4$ with those at $\Nf=2+1$,
assuming the chiral limit for the light quarks and a physical strange quark mass. The $\mu$-$\Nf$ is limited by the line $\mu_c(T=0)$, beyond which chiral symmetry is restored. \emph{Left:} 
For $\Nf=4$, the transition is of first order for all values of $\mu$.
\emph{Center:} The tricritical point at $\Nf=4$ is is connected to the tricritical point at $\Nf=2+1$. This would be evidence for the existence of the
critical end point in the QCD phase diagram for physical quark masses.
\emph{Right:} The $\Nf=4$ first order region does not extend to $\Nf=2+1$, where it remains second order.
This second order transition turns into a crossover immediately as $m_u,m_d>0$, so in this scenario there is no chiral critical end point at physical quark masses.
}
\label{Scenarios}
\begin{minipage}{0.35\textwidth}
\includegraphics[width=\textwidth]{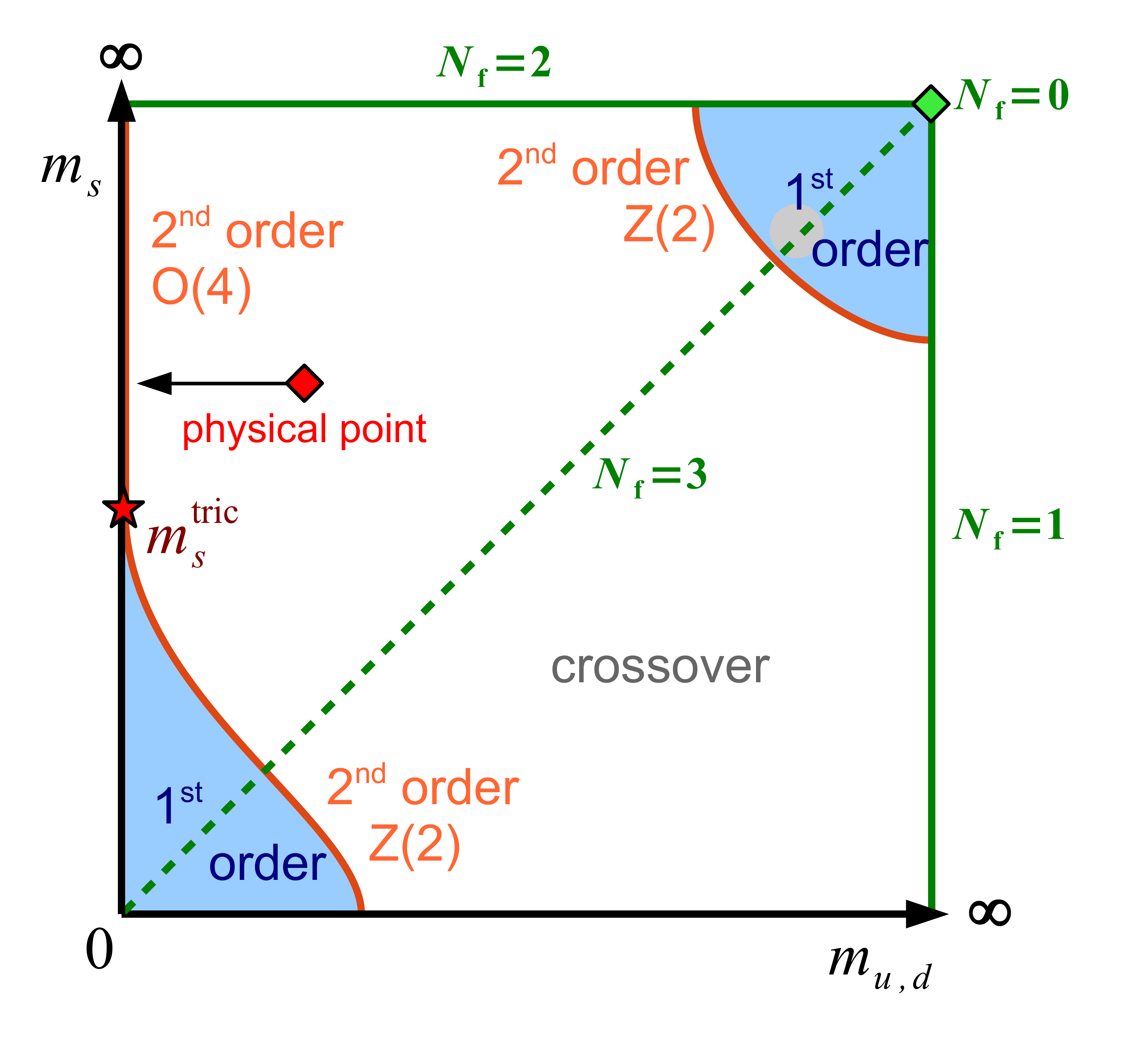}
\end{minipage}
\begin{minipage}{0.64\textwidth}
\caption{The Columbia plot with the assumption $m_s^{phys}>m_s^{tric}$, which implies that the chiral transition is second order for $\Nf=2$.
The arrow points towards the $\Nf=2+1$ chiral light quark masses and physical strange quark mass as denoted in the bottom row of Fig.~4 in between $\Nf=2$ and $\Nf=3$.
\vspace{15mm}
\label{Columbia}
}
\end{minipage}
\end{figure}

In Fig.~\ref{Scenarios} we speculate how the separation of the first order transitions could be realized at larger values of $\beta$. 
Moreover, we can distinguish at least three scenarios (A,B,C) on how the chiral tricritical point depends on $\beta$. 
These scenarios start from the same phase diagram in the strong coupling limit, but have different continuum limits at $\beta\rightarrow \infty$ ($a\rightarrow 0$).
In all three scenarios, a tricritical point exists at $\mu=0$, $\beta>0$: it must exist because the finite-temperature $\mu=0$ transition, which is of second order for $\beta=0$,
is of first order for $\beta=\infty$, following the argument of \cite{Pisarski1983} which applies to the continuum, $\Nf=4$ theory.
\begin{enumerate}
\item In scenario (A) the chiral transition remains first order for all values of $\mu_B$. Hence the tricritical line turns towards $\mu=0$ at some finite $\beta_{\rm tric}^{(\mu=0)}$.
\item In scenario (B) the chiral transition weakens and hence turns second order, but strengthens again to turns first order at larger $\mu_B$. 
\item In scenario (C) the chiral transition weakens and remains second order. In that case the tricritical line bends towards larger $\mu$ and eventually vanishes at some finite $\beta_{tric}^{(T=0)}$.
\end{enumerate}

In order to discuss the relation between the phase diagram in the $\mu$-$T$ plane for $\Nf=4$ massless quarks with the more physical scenario $\Nf=2+1$ with 2 massless up and down quarks and one physical strange quark, 
we show phase diagrams in the $\Nf$-$\mu$ plane. Interpolating between integer numbers of massless flavors $\Nf$ and $\Nf+1$ can be realized by decreasing the mass of an additional flavor from infinity to zero.
In all scenarios it is assumed that for $\Nf=2$, the chiral transition is second order, and that there is a tricritical strange quark mass $m_s^{tric}$ separating it from the 
$\Nf=3$ first order transition, as shown in the 
so-called Columbia plot, Fig.~\ref{Columbia}.
Note that whether $\Nf=2$ is indeed second order and thus whether $m_s^{tric}$ exists and also whether it is larger or smaller than the physical strange quark mass is still under debate \cite{Bonati2014}.
The standard scenario of QCD in the chiral limit, as shown in Fig.~\ref{PhaseDiagCL} (right), corresponds to scenario (B) in Fig \ref{Scenarios}. 
However, the non-standard scenario (C) is supported by Monte Carlo simulations for imaginary chemical potential and analytic continuation \cite{Bonati2014,Forcrand2008}: these studies suggest 
(at least for small chemical potential) that the chiral transition weakens with chemical potential, making the $\Nf=3$ first order region in Fig.~\ref{Columbia} to shrink with increasing $\mu_B$. This should also be the case for $\Nf=4$.

A last comment on staggered fermions is in order: one of the lattice artifacts is due to the way this discretization solves the so-called fermion doubling problem:
At strong coupling, there is effectively only one quark flavor, whereas in the continuum limit the same action describes 4 flavors due to the fermion doubling. 
Instead of 15 Goldstone bosons that are present in the $\Nf=4$ continuum theory, there is only one Goldstone boson at strong coupling, since the others 14 receive masses from lattice artifacts (called taste-splitting).
In the determinant-based approach, the problem is solved by ``rooting'': taking the root of the fermion determinant to reduce the number of flavors from 4 to 2 (and the number of Goldstone bosons from 15 to 3).
This strategy is not available in our dual-variable approach. 
Although the strong coupling limit has effectively only one flavor, the residual chiral symmetry is that of a $\Nf=4$ continuum theory, with one true Goldstone boson which even persists when the chiral anomaly
$U_{A}(1)$ is present for $\beta>0$. This is in contrast to a genuine $\Nf=1$ theory in the continuum which has no Goldstone bosons at all. The chiral anomaly breaks the chiral symmetry explicitly, driving the chiral transition into a 
crossover (corresponding to the lower right corner of the Columbia plot Fig.~\ref{Columbia}). Hence the deconfinement transition at $\Nf=0$ is most likely completely separate from the chiral transition for $\Nf\geq 2$, as shown in
all three scenarios Fig.~\ref{Scenarios} (bottom). 

\section{Outlook for future investigations}

There are various ways to discretize fermions on the lattice, with staggered fermions and Wilson fermions the most widely used for thermodynamics studies.
They describe the same physics in the continuum limit only. At finite lattice spacing, and in particular at strong coupling, both discretizations are quite different.
In particular, the spin and the kinetic term of the fermion action are treated very differently. 
A dimer+flux representation is also possible for Wilson fermions. Such a representation was so far only determined for lattice QED \cite{Salmhofer1991,Scharnhorst1996}, since the Grassmann integration is much more involved for $N_c>1$.

As a matter of principle, for both lattice discretizations, the gauge action can be incorporated order by order in $\beta$. There are however technical difficulties that remain to be solved.
A new strategy to study both lattice discretizations on a par is to expand both in systematically in $\beta$ and the inverse quark mass by making use of a Hamiltonian formulation \cite{Unger2012}. The partition function is then expressed by a Hamiltonian composed by operators:
\begin{align}
Z&=\Tr[e^{\beta \mathcal{H}}],& \mathcal{H}&=
\frac{1}{2}\sum_{\langle x,y \rangle}
\sum_{Q_i}J^+_{Q_i(x)} J^-_{Q_i(y)},&
J^{-}_{Q_i}&=(J^{+}_{Q_i})^\dagger
\end{align}
where the generalized quantum numbers $Q_i$ (spin, parity,flavor) are globally conserved, and nearest neighbor interactions are characterized by the operators
$J^+_{Q_i(x)} J^-_{Q_i(y)}$, which raise the quantum number $Q_i$ at site $x$ and lowers it at a neighboring site $y$ (see \cite{Unger2012} for the case of 
$\Nf=1,2$ for staggered fermions). For both staggered fermions and Wilson fermions, the matrices $J^{\pm}_{Q_i}$ contain vertex weights which are the crucial input to sample 
the corresponding partition function. The plan for the future is to do so with a quantum Monte Carlo algorithm.
Comparing both fermion discretizations order by order in the strong coupling expansion will help to discriminate lattice discretization errors from the genuine physics, in particular with respect to QCD at finite density.\\

\emph{Acknowledgement} - This works was supported by the Helmholtz International Center for FAIR
within the LOEWE program launched by the State
of Hesse.


\begin{thebibliography}{99}

\bibitem{Kawamoto1981}
  N.~Kawamoto and J.~Smit,
  Nucl.\ Phys.\ B {\bf 192} (1981) 100.

  \bibitem{Damgaard1985}
  P.~H.~Damgaard, D.~Hochberg and N.~Kawamoto,
  Phys.\ Lett.\ B {\bf 158} (1985) 239.

  \bibitem{Bilic1992a}
  N.~Bilic, F.~Karsch, K.~Redlich,
  {\em Phys. Rev. D {\bf 45}} (1992) 3228.


\bibitem{Bilic1992b}
  N.~Bilic, K.~Demeterfi, B.~Petersson,
  {\em Nucl. Phys. B {\bf 377}} (1992) 3651.


\bibitem{Nishida2004}
Y.~Nishida,
{\em Phys. Rev. D {\bf 69}} (2004) 094501.

\bibitem{Ohnishi2009}
  K.~Miura, T.~Z.~Nakano, A.~Ohnishi and N.~Kawamoto,
  {\em Phys.\ Rev.\ D {\bf 80}} (2009) 074034.

  
\bibitem{Rossi1984}
P.~Rossi, U.~Wolff,
{\em Nucl. Phys. B {\bf 258}} (1984) 105;\quad 

   \bibitem{Wolff1985}
U.~Wolff,
{\em Phys. Lett. B {\bf 153}} (1985) 92.

  \bibitem{Karsch1989}
F.~Karsch, K.~H.~M\"utter,
{\em Nucl. Phys. B {\bf 313}} (1989) 541.
  
\bibitem{Adams2003}
  D.~H.~Adams and S.~Chandrasekharan,
  {\em Nucl.\ Phys.\  B {\bf 662}} (2003) 220.

\bibitem{Fromm2010}
  P.~de Forcrand and M.~Fromm,
  Phys.\ Rev.\ Lett.\  {\bf 104} (2010) 112005
  [arXiv:0907.1915 [hep-lat]].
  

  
  \bibitem{Unger2011}
  W.~Unger and P.~de Forcrand,
  PoS LATTICE {\bf 2011} (2011) 218
  [arXiv:1111.1434 [hep-lat]].
  
\bibitem{Unger2014}
P.~de Forcrand, J.~Langelage, O.~Philipsen and W.~Unger,
{\em Phys. Rev. Lett. {\bf 113}} (2014) 152002.

\bibitem{Pisarski1983}
  R.~D.~Pisarski and F.~Wilczek,
  Phys.\ Rev.\ D {\bf 29} (1984) 338.

  \bibitem{Bonati2014}
  C.~Bonati, P.~de Forcrand, M.~D'Elia, O.~Philipsen and F.~Sanfilippo,
  Phys.\ Rev.\ D {\bf 90} (2014) 7,  074030
  [arXiv:1408.5086 [hep-lat]].

  \bibitem{Forcrand2008}
  P.~de Forcrand and O.~Philipsen,
  JHEP {\bf 0811} (2008) 012
  [arXiv:0808.1096 [hep-lat]].

\bibitem{Salmhofer1991}
  M.~Salmhofer,
  Nucl.\ Phys.\ B {\bf 362} (1991) 641.

\bibitem{Scharnhorst1996}
  K.~Scharnhorst,
  Nucl.\ Phys.\ B {\bf 479} (1996) 727
  [hep-lat/9604024].
  
\bibitem{Unger2012}
  W.~Unger and P.~de Forcrand,
  PoS LATTICE {\bf 2012} (2012) 194
  [arXiv:1211.7322 [hep-lat]].
  


%
%
%
%
%
%
%
%
%

  
\end{thebibliography}
\end{document}